\input harvmac
\input tables
\input amssym.tex
\input amssym.def

\font\teneurm=eurm10 \font\seveneurm=eurm7 \font\fiveeurm=eurm5
\newfam\eurmfam
\textfont\eurmfam=\teneurm \scriptfont\eurmfam=\seveneurm
\scriptscriptfont\eurmfam=\fiveeurm

 \font\teneusm=eusm10 \font\seveneusm=eusm7 \font\fiveeusm=eusm5
\newfam\eusmfam
\textfont\eusmfam=\teneusm \scriptfont\eusmfam=\seveneusm
\scriptscriptfont\eusmfam=\fiveeusm

\font\tencmmib=cmmib10 \skewchar\tencmmib='177
\font\sevencmmib=cmmib7 \skewchar\sevencmmib='177
\font\fivecmmib=cmmib5 \skewchar\fivecmmib='177
\font\ninett=cmtt10   scaled 900  
\newfam\cmmibfam
\textfont\cmmibfam=\tencmmib \scriptfont\cmmibfam=\sevencmmib
\scriptscriptfont\cmmibfam=\fivecmmib

\writedefs

\noblackbox\input rotate
\def\figin{\epsfcheck\figin}\def\figins{\epsfcheck\figins}
\def\epsfcheck{\ifx\epsfbox\UnDeFiNeD
\message{(NO epsf.tex, FIGURES WILL BE IGNORED)}
\gdef\figin##1{\vskip2in}\gdef\figins##1{\hskip.5in}
\else\message{(FIGURES WILL BE INCLUDED)}%
\gdef\figin##1{##1}\gdef\figins##1{##1}\fi}
\def\DefWarn#1{}

\def\cG{{\cal G}}
\def\cH{{\cal H}}

\def\cO{{\cal O}}

\def\cZ{{\cal Z}}

\def\bbZ{{\Bbb Z}}

\def\D{\Delta}
\def\e{\epsilon}

\def\n{\nu}

\def\om{\omega}
\def\Om{\Omega}

\def\ii{{\rm i}}

\def\Klsum{\mathop{\rm Kl}}

\noblackbox

\Title{\vbox{\baselineskip12pt\hbox{\hss 0708.3386 [hep-th]}}}
{\vbox{ \centerline{Three-dimensional AdS gravity and}
\bigskip
\centerline{extremal CFTs at $c=8 m$}}}
\smallskip
\centerline{Spyros D. Avramis${}^{1,2}$, Alex Kehagias${}^{2}$ and
Constantina Mattheopoulou${}^{2}$} \smallskip
\centerline{${}^{1}$\it{Department of Engineering Sciences,
University of Patras,}} \centerline{\it{26110 Patras, Greece}}
\smallskip
\centerline{${}^{2}$\it{Physics Department, National Technical
University of Athens,}} \centerline{\it{15780 Zografou Campus,
Athens, Greece}}
\smallskip
{\ninett avramis@mail.cern.ch, kehagias@central.ntua.gr,
conmat@central.ntua.gr}
\bigskip\bigskip

\medskip
\noindent We note that Witten's proposed duality between extremal
$c=24 k$ CFTs and three-dimensional anti-de Sitter gravity may
possibly be extended to central charges that are multiples of $8$,
for which extremal self-dual CFTs are known to exist up to $c=40$.
All CFTs of this type with central charges $c \geqslant 24$,
provided that they exist, have the required mass gap and may serve
as candidate duals to three-dimensional gravity at the
corresponding values of the cosmological constant. Here, we
compute the genus one partition function of these theories up to
$c=88$, we give exact and approximate formulas for the
degeneracies of states, and we determine the genus two partition
functions of the theories up to $c=40$.

\Date{August 2007}

\lref\GottQG{
  J.R.~Gott and M.~Alpert,
  ``General Relativity In A (2+1)-Dimensional Space-Time,''
  Gen.\ Rel.\ Grav.\  {\bf 16}, 243 (1984).
}

\lref\BrownAM{
  J.D.~Brown,
  {\it Lower dimensional gravity},
  World Scientific (1988).
}

\lref\CarlipUC{
  S.~Carlip,
  {\it Quantum gravity in 2+1 dimensions},
  Cambridge University Press (1998).
}

\lref\MartinecFS{
  E.J.~Martinec,
  ``Soluble Systems in Quantum Gravity,''
  Phys.\ Rev.\  D {\bf 30}, 1198 (1984).
}

\lref\WittenHC{
  E.~Witten,
  ``(2+1)-Dimensional Gravity as an Exactly Soluble System,''
  Nucl.\ Phys.\  B {\bf 311}, 46 (1988).
}

\lref\DeserTN{
  S.~Deser, R.~Jackiw and G.~'t Hooft,
  ``Three-Dimensional Einstein Gravity: Dynamics of Flat Space,''
  Annals Phys.\  {\bf 152}, 220 (1984).
}

\lref\DeserDR{
  S.~Deser and R.~Jackiw,
  ``Three-Dimensional Cosmological Gravity: Dynamics of Constant Curvature,''
  Annals Phys.\  {\bf 153}, 405 (1984).
}

\lref\WittenKT{
  E.~Witten,
  ``Three-Dimensional Gravity Revisited,''
  {\tt arXiv:0706.3359}.
}

\lref\AchucarroVZ{
  A.~Ach\'ucarro and P.K.~Townsend,
  ``A Chern-Simons Action for Three-Dimensional anti-De Sitter Supergravity
  Theories,''
  Phys.\ Lett.\  B {\bf 180}, 89 (1986);
  ``Extended Supergravities in d=(2+1) as Chern-Simons Theories,''
  Phys.\ Lett.\  B {\bf 229}, 383 (1989).
}

\lref\CarlipGY{
  S.~Carlip,
  ``The Statistical Mechanics Of The (2+1)-Dimensional Black Hole,''
  Phys.\ Rev.\  D {\bf 51}, 632 (1995),
  {\tt gr-qc/9409052}.
}

\lref\BrownNW{
  J.D.~Brown and M.~Henneaux,
  ``Central Charges in the Canonical Realization of Asymptotic Symmetries: An
  Example from Three-Dimensional Gravity,''
  Commun.\ Math.\ Phys.\  {\bf 104}, 207 (1986).
}

\lref\BanadosWN{
  M.~Ba\~nados, C.~Teitelboim and J.~Zanelli,
  ``The Black Hole in Three-Dimensional Space-time,''
  Phys.\ Rev.\ Lett.\  {\bf 69}, 1849 (1992),
  {\tt hep-th/9204099}.
}

\lref\BanadosGQ{
  M.~Ba\~nados, M.~Henneaux, C.~Teitelboim and J.~Zanelli,
  ``Geometry of the (2+1) Black Hole,''
  Phys.\ Rev.\  D {\bf 48}, 1506 (1993),
  {\tt gr-qc/9302012}.
}

\lref\CarlipZN{
  S.~Carlip,
  ``Conformal Field Theory, (2+1)-Dimensional Gravity, and the BTZ Black
  Hole,''
  Class.\ Quant.\ Grav.\  {\bf 22}, R85 (2005),
  {\tt gr-qc/0503022}.
}

\lref\CarlipGY{
  S.~Carlip,
  ``The Statistical Mechanics of the (2+1)-Dimensional Black Hole,''
  Phys.\ Rev.\  D {\bf 51}, 632 (1995),
  {\tt gr-qc/9409052}.
}

\lref\StromingerEQ{
  A.~Strominger,
  ``Black Hole Entropy from Near-horizon Microstates,''
  JHEP {\bf 9802}, 009 (1998),
  {\tt hep-th/9712251}.
}

\lref\VerlindeSN{
  E.P.~Verlinde,
  ``Fusion Rules and Modular Transformations in 2d Conformal Field Theory,''
  Nucl.\ Phys.\  B {\bf 300}, 360 (1988).
}

\lref\Hoehn{
  G. H\"ohn,
  ``{\it Selbstduale Vertexoperatorsuperalgebren und das Babymonster},''
  Ph.D. thesis  (Bonn 1995),
  Bonner Mathematische Schriften {\bf 286}, 1-85 (1996),
  {\tt arXiv:0706.0236}.
}

\lref\SchellekensDB{
  A.N.~Schellekens,
  ``Meromorphic $c=24$ Conformal Field Theories,''
  Commun.\ Math.\ Phys.\  {\bf 153}, 159 (1993),
  {\tt hep-th/9205072}.
}

\lref\FLM{
  I.~Frenkel, J.~Lepowsky and A.~Meurman,
  ``A Natural Representation of the Fischer-Griess Monster with the Modular Function $J$ as
  Character,'' Proc.\ Natl.\ Acad.\ Sci.\ USA\  {\bf 81} (10), 3256
  (1984);
  {\it Vertex operator algebras and the monster},
  Academic Press, New York (1988).
}

\lref\DixonQD{
  L.J.~Dixon, P.H.~Ginsparg and J.A.~Harvey,
  ``Beauty and the Beast: Superconformal Symmetry in a Monster Module,''
  Commun.\ Math.\ Phys.\  {\bf 119}, 221 (1988).
}

\lref\ManschotZB{
  J.~Manschot,
  ``${\rm AdS}_3$ Partition Functions Reconstructed,''
  {\tt arXiv:0707.1159}.
}

\lref\Petersson{
  H.~Petersson,
  ``Uber die Entwicklungskoeffizienten der Automorphen Formen,''
  Acta.\ Math.\  {\bf 58}, 169 (1932).
}

\lref\Rademacher{
  H.~Rademacher,
  ``The Fourier Coefficients of the Modular Invariant $J(\tau)$,''
  Amer.\ J.\ Math.\  {\bf 60}, 501 (1938).
}

\lref\DijkgraafFQ{
  R.~Dijkgraaf, J.M.~Maldacena, G.W.~Moore and E.P.~Verlinde,
  ``A Black Hole Farey Tail,''
  {\tt hep-th/0005003}.
}

\lref\BirminghamXD{
  D.~Birmingham and S.~Sen,
  ``An Exact Black Hole Entropy Bound,''
  Phys.\ Rev.\  D {\bf 63}, 047501 (2001),
  {\tt hep-th/0008051};

  D.~Birmingham, I.~Sachs and S.~Sen,
  ``Exact Results for the BTZ Black Hole,''
  Int.\ J.\ Mod.\ Phys.\  D {\bf 10}, 833 (2001),
  {\tt hep-th/0102155}.
}

\lref\KaulKF{
  R.K.~Kaul and P.~Majumdar,
  ``Logarithmic Correction to the Bekenstein-Hawking entropy,''
  Phys.\ Rev.\ Lett.\  {\bf 84}, 5255 (2000),
  {\tt gr-qc/0002040};

  T.R.~Govindarajan, R.K.~Kaul and V.~Suneeta,
  ``Logarithmic Correction to the Bekenstein-Hawking Entropy of the BTZ Black
  Hole,''
  Class.\ Quant.\ Grav.\  {\bf 18}, 2877 (2001)
  {\tt gr-qc/0104010};

}

\lref\CardyIE{
  J.L.~Cardy,
  ``Operator Content Of Two-Dimensional Conformally Invariant Theories,''
  Nucl.\ Phys.\  B {\bf 270}, 186 (1986).
}

\lref\GaiottoXH{
  D.~Gaiotto and X.~Yin,
  ``Genus Two Partition Functions of Extremal Conformal Field Theories,''
  {\tt arXiv:0707.3437}.
}

\lref\Freitag{
  E.~Freitag,
  {\it Siegelsche modulformen},
  Grundlehren der mathematischen Wissenschaften, Bd. {\bf 254},
  Springer-Verlag (1983).
}

\lref\Klingen{
  H.~Klingen,
  {\it Introductory lectures on Siegel modular forms},
  Cambridge University Press (1990).
}

\lref\Igusa{
  J.~Igusa,
  ``On Siegel Modular Forms of Genus Two,''
  Amer.\ J.\ Math.\  {\bf 84} 175-200 (1962);\
  ``On Siegel Modular Forms of Genus Two (II),''
  Amer.\ J.\ Math.\ {\bf 86}, 392-412 (1964).
}

\lref\TuiteID{
  M.P.~Tuite,
  ``Genus Two Meromorphic Conformal Field Theory,''
  {\tt math/9910136}.
}

\lref\MasonDK{
  G.~Mason and M.P.~Tuite,
  ``On Genus Two Riemann Surfaces Formed from Sewn Tori,''
  Commun.\ Math.\ Phys.\  {\bf 270}, 587 (2007),
  {\tt math/0603088}.
}

\lref\SonodaMF{
  H.~Sonoda,
  ``Sewing Conformal Field Theories,''
  Nucl.\ Phys.\  B {\bf 311}, 401 (1988);
  ``Sewing Conformal Field Theories. 2,''
  Nucl.\ Phys.\  B {\bf 311}, 417 (1988).
}

\lref\Zhu{Y.~Zhu,
  ``Modular Invariance of Characters of Vertex Operator Algebras,''
    J.\ Amer.\ Math.\ Soc.\  {\bf 9}, 237–307 (1996).
}

\lref\DongEA{
  C.Y.~Dong, H.S.~Li and G.~Mason,
  ``Modular-Invariance of Trace Functions in Orbifold Theory,''
  Commun.\ Math.\ Phys.\  {\bf 214}, 1 (2000),
  {\tt q-alg/9703016}.
}

\lref\GaberdielVE{
  M.R.~Gaberdiel,
  ``Constraints on Extremal Self-Dual CFTs,''
  {\tt arXiv:0707.4073}.
}

\newsec{Introduction}

Three-dimensional gravity \refs{\GottQG,\BrownAM,\CarlipUC} is a
promising candidate for a potentially soluble
\refs{\MartinecFS,\WittenHC} gravitational theory with non-trivial
structure. In this respect, the most interesting case is that of
negative cosmological constant, where the AdS/CFT correspondence
maps the problem of solving the theory to the more concrete one of
identifying the dual CFT, and where the theory admits, in addition
to the standard conical solutions \refs{\DeserTN,\DeserDR}, the
BTZ black hole solutions \refs{\BanadosWN,\BanadosGQ}. The latter
share many properties with their four-dimensional counterparts
and, in particular, have a nonzero entropy which should be
accounted for by the microstates of any candidate dual CFT. These
considerations led Witten \WittenKT\ to re-examine the theory and
to propose a duality with a certain class of self-dual CFTs.
Below, we outline the salient points of the proposal.

An important property of three-dimensional gravity is that it
admits a description in terms of a Chern-Simons theory
\AchucarroVZ. For ${\rm AdS}_3$ gravity with cosmological constant
$\Lambda=-1/\ell^2$, plus a possible gravitational Chern-Simons
term, the corresponding action,
\eqn\gravityaction{S = {1 \over 16 \pi G} \int d^3 x \sqrt{-g} \left( R + {2 \over \ell^2} \right) +
{k' \over 4\pi} \int \Tr \left( \om \wedge d \om + {2 \over 3} \om
\wedge \om \wedge \om \right)\ ,
}
can be recast \WittenHC\ into the following combination of
Chern-Simons actions
\eqn\csaction{\eqalign{&\qquad S = S_L - S_R\ ,
\cr S_{L,R} = {k_{L,R} \over 4\pi} \int \Tr & \left( A_{L,R} \wedge
d A_{L,R} + {2 \over 3} A_{L,R} \wedge A_{L,R} \wedge A_{L,R}
\right)\ .}
}
Here, $A_{L,R}^a = \om^a \mp {1 \over \ell} e^a$ are the $SO(2,1)
\times SO(2,1)$ gauge fields, expressed in terms of the dreibein
$e^a$ and the dual spin connection $\om^a = {1 \over 2} \e^{abc}
\om_{bc}$, while $k_{L,R} = {\ell \over 16 G} \pm {k' \over 2}$.
For pure AdS gravity, where $k'=0$ and $k_L = k_R = k = \ell /16
G$, it was first shown by Brown and Henneaux \BrownNW\ that the
Poisson-bracket algebra of the asymptotic symmetries consists of
two copies of the Virasoro algebra with common central charge $c=3
\ell /2G = 24 k$; in modern terminology this is understood as the
central charge of the dual CFT. For $k' \ne 0$,
$c_{L,R}=24k_{L,R}$ are different from each other.

Regarding the quantum theory, the Chern-Simons formulation has the
conceptual advantages that it makes finiteness of
three-dimensional gravity manifest and provides a good starting
point for formulating perturbation theory. However, turning to
nonperturbative aspects, there are a few problems, the most
important of which being that the Chern-Simons theory appears to
have too few degrees of freedom to account for the degeneracy of
BTZ black holes (unless one associates them with boundary
excitations \CarlipGY). As the above make clear that this
formulation must be used with caution, the approach advocated by
Witten in \WittenKT\ was to use the Chern-Simons description only
as a guide for obtaining the relevant values of the central
charges of the boundary CFT, and then to try determining the
latter by imposing modular invariance and the existence of a mass
gap.

The relevant values of $c_{L,R}$ are obtained from the
quantization conditions on the Chern-Simons couplings $k_{L,R}$.
To find these conditions, it must be specified whether the gauge
group is precisely $SO(2,1) \times SO(2,1)$ or a cover thereof.
Specifically, for an $n$-fold diagonal cover of $SO(2,1) \times
SO(2,1)$, the quantization conditions are given by \WittenKT
\eqn\quantk{k_L \in {1 \over n} \bbZ\ (\hbox{$n$ odd})\ , \, {1 \over 2n} \bbZ\ (\hbox{$n$
even})\ ,\qquad k' = k_L - k_R \in \bbZ\ \left(\hbox{or }{1 \over
3} \bbZ\right)\ ,
}
where the relaxed condition $k' \in {1 \over 3} \bbZ$ arises when
$\om$ is treated like a connection on the tangent bundle rather
than like an ordinary gauge field. In the simplest $SO(2,1) \times
SO(2,1)$ case, $k_{L,R}$ are integers and $c_{L,R}=24 k_{L,R}$ are
multiples of $24$; for pure gravity this means that $k = \ell / 16
G$ is an integer. We stress that there is no a priori reason for
picking a particular value of $n$; the only restriction is that we
cannot take $n \to \infty$, as this would allow us to continuously
vary $c = 3 \ell / 2 G$ in contradiction with Zamolodchikov's
$c$-theorem.

The cases $c=24 k$, $k \in \bbZ$, are rather special in that they
are the only ones that allow for the possibility of holomorphic
factorization. In the self-dual case, where the space of states of
the theory consists only of the vacuum representation, holomorphic
factorization implies that the partition function factorizes as
\eqn\holfac{\cZ_c(\tau,\bar{\tau}) = Z_c(\tau)
\bar{Z}_c(\bar{\tau})\ ,
}
with $Z_c(\tau)$ and $\bar{Z}_c(\bar{\tau})$ being separately
modular invariant. Although there exists no compelling argument
for considering only the cases $c=24 k$, the decomposition
\csaction\ of the action as a sum of two terms is suggestive of a
factorization of the form \holfac, and the absence of CFTs with
the required properties at lower central charges points towards
the value $c=24$ above which there is a proliferation of
holomorphic CFTs. Be that as it may, the assumption of holomorphic
factorization tremendously simplifies things, as one may uniquely
determine the partition functions of the sought CFTs by imposing
modular invariance and requiring that the primaries associated
with black-hole states enter at the highest possible level. For
$c=24$ and $c=48$, the (holomorphic) partition functions read
\eqn\monst{\eqalign{Z_{24} (\tau) &= j (\tau) - 744 \cr &= q^{-1}
+ 196884\, q + 21493760\, q^2 + 864299970\, q^3 + 20245856256\,
q^4 + \ldots\ , } }
and
\eqn\next{\eqalign{Z_{48} (\tau) &=
j^2(\tau) - 1488\, j(\tau) + 159769 \cr &= q^{-2} + 1 + 42987520\,
q + 40491909396\, q^2 + 8504046600192\, q^3 + \ldots\ ,}
}
where $j(\tau)$ is the modular $j$-function and $q=e^{2\pi \ii
\tau}$. The partition function in \monst\ defines a very special
theory among the 71 holomorphic CFTs believed to exist at $c=24$
\SchellekensDB. It was first constructed by Frenkel, Lepowsky and
Meurman \FLM\ (see also \DixonQD) by considering 24 chiral bosons
on the Leech lattice and using a $\bbZ_2$ orbifold to project out
the 24 dimension-1 primaries. The 196884 dimension-2 operators
correspond to one Virasoro descendant plus 196883 primaries whose
number is the dimension of the lowest non-trivial representation
of the largest sporadic group, the monster group. In fact, each
coefficient in \monst\ equals the number of descendants at this
level plus the dimension of an irreducible representation of the
monster; this observation forms part of monstrous moonshine, an
unexpected connection between modular functions and finite simple
groups. For the partition function \next\ and, in general, all
$Z_{24k}(\tau)$ with $k \geqslant 2$, the corresponding CFTs have
not been identified, but the number of available lattices in these
dimensions makes their existence plausible. Furthermore, the
counting of microstates in the CFTs under consideration yields,
for all values of $k$, an entropy that is very close to the
corresponding Bekenstein-Hawking entropy of the BTZ black hole.
These facts led Witten to propose that three-dimensional quantum
gravity with $\ell / 16 G = k \in \bbZ$ is dual to the $c=24k$
series of extremal CFTs. In particular, for the most negative
possible value of the cosmological constant, the dual CFT has been
conjectured to be the $c=24$ monster theory.

\newsec{Genus one extremal partition functions for $c=8m$}

The main purpose of this paper is to note that the arguments
mentioned above may apply, with minor modifications, to the case
where the central charge is a multiple of 8, $c=8 m$ with $m \in
\bbZ$. In the Chern-Simons formulation this corresponds to taking
a three-fold diagonal cover of $SO(2,1) \times SO(2,1)$ as the
gauge group. Note that the resulting values of the Chern-Simons
couplings, $k_{L,R} \in {1 \over 3} \bbZ$, fit nicely with the
last quantization condition in \quantk. In the cases $c=8 m$,
holomorphic factorization is no longer possible in the strict
sense, but one can still have holomorphic factorization up to a
phase. This, along with the requirement that the primaries
associated with black holes appear at the right level, uniquely
specifies the partition function for each value of $m$. In what
follows, we describe the construction of these partition
functions, and we state exact and approximate formulas for the
corresponding degeneracies of states.

\subsec{Partition functions}

The starting point of our construction is the well-known fact
that, for a CFT of central charge $c=8m$, there exists the
possibility that the partition function factorizes as in \holfac,
but with the holomorphic part picking up a phase under $T \, : \,
\tau \to \tau+1$,
\eqn\phase{ Z_{8 m} (\tau) \to  e^{- 2 \pi \ii m / 3} Z_{8m} (\tau) \
,
}
and with the antiholomorphic part picking up the opposite phase so
that the full partition function is modular invariant (see e.g.
\refs{\SchellekensDB,\VerlindeSN}). Assuming that this is the case
and furthermore assuming as in \WittenKT\ that the theory is
self-dual, the construction proceeds as follows. In the absence of
primary fields, the partition function of such a theory would be
just the vacuum Virasoro character,
\eqn\znaive{ Z_{0,8m} (\tau) = q^{- m / 3} \prod_{n=2}^{\infty} {1 \over 1-q^n}
= q^{- m / 3} \sum_{n=0}^\infty \left( P(n)- P(n-1) \right) q^n\ ,
}
where $P(n)$ denotes the number of partitions of $n$. This
partition function clearly cannot account for the degeneracy of
the BTZ black holes and, in addition, transforms non-trivially
under $S \, : \, \tau \to - 1 / \tau$. To remedy these problems we
would like to add primaries, to be identified with operators
creating BTZ black holes, in such a way that modular invariance up
to a phase is restored. To figure out the conformal weight of such
states, we note that, choosing the additive constant in $L_0$ so
that its eigenvalue is $h-{c \over 24}$ where $h$ is the conformal
weight, $L_0$ is related to the mass and angular momentum of a BTZ
black hole according to
\eqn\abcd{ L_0 = {1 \over 2} (\ell M + J)\ ,
}
and the Bekenstein-Hawking entropy reads \StromingerEQ
\eqn\SBH{ S_{BH} (m,L_0) = 4 \pi \sqrt{{c \over 24} L_0} = 4 \pi \sqrt{{m \over 3} L_0}\ ,
}
with similar relations for the antiholomorphic sector. The minimal
mass of a black hole corresponds to the case $\ell M = | J |$,
i.e. $L_0=0$, for which the entropy vanishes. Therefore, the
primaries associated with the black-hole states should appear for
$L_0 > 0$, i.e. for $h \geqslant h_m$ where
\eqn\abcd{h_m \equiv \left[ {m \over 3} \right] + 1\ .
}
On the other hand, according to a result of H\"ohn \Hoehn, the
dimension of the lowest primary in a self-dual CFT has an upper
limit given by $h \leqslant h_m$.  Therefore, our requirements can
be satisfied only if $h = h_m$, that is, if the full partition
function has the form
\eqn\zfull{ Z_{8m} (\tau) = q^{- m / 3} \left( \prod_{n=2}^{\infty}
{1 \over 1-q^n} + \cO (q^{\left[ m / 3 \right] + 1}) \right) \ .
}
Such partition functions are called extremal and have the
remarkable property that they are uniquely determined once one
imposes modular invariance up to a phase. Namely, the requirement
\phase\ fixes a self-dual partition function with $c=8m$ to be a
weighted polynomial of weight $m/3$ generated by $j^{1/3}(\tau)$,
with the general form \Hoehn\
\eqn\zansatz{ Z_{8m} (\tau) = j^{m/3} (\tau)
\sum_{r=0}^{\left[ m/3 \right]} a_r j^{-r} (\tau) \ .
}
The coefficients $a_r$ are then determined by matching the terms
of order $q^{r- m / 3}$, $r=0,\ldots,\left[ m / 3 \right]$, with
those in \zfull\ as in \WittenKT\ (see also \ManschotZB). The
results of this analysis are given below.

For $c=8, \, 16$, we have $h_m=1$ meaning that the extra states
enter at level one above the vacuum. Therefore there exists no
mass gap and the extra states must correspond to massless fields
arising as a result of a gauge symmetry. In fact, as the self-dual
partition functions for $c=8$ and $c=16$ are well-known and
believed to be unique, the result can be immediately anticipated.
Indeed, applying the matching procedure we find
\eqn\zeight{Z_{8} (\tau) = j^{1/3} (\tau) = q^{-1/3} +
248\, q^{2/3} + 4124\, q^{5/3} + 34752\, q^{8/3} + 213126\,
q^{11/3} + \ldots\ ,
}
which is the vacuum character of the level 1 affine $\hat{E}_8$
theory (or $q^{-1/3}$ times the McKay-Thompson series of class 3C
for the monster) and
\eqn\zsixteen{\eqalign{Z_{16} (\tau) &= j^{2/3} (\tau) \cr &= q^{-2/3} +
496\, q^{1/3} + 69752\, q^{4/3} + 2115008\, q^{7/3} + 34670620\,
q^{10/3} + \ldots \, }
}
which is the vacuum character of the level 1 affine $\hat{E}_8
\times \hat{E}_8$ theory. Due to the presence of Kac-Moody
symmetries, these extremal CFTs cannot be directly relevant to
pure ${\rm AdS_3}$ gravity but, possibly, to extensions of ${\rm
AdS_3}$ gravity including gauge fields with Chern-Simons
interactions \WittenKT.

For $c=24,\, 32,\, 40$, we have $h_m=2$, i.e. the extra states
enter at level two above the vacuum and the required mass gap
exists. For $c=24$ we find the partition function \monst\
discussed earlier on. For $c=32$ and $c=40$, we obtain the
partition functions
\eqn\zthirtytwo{\eqalign{Z_{32} (\tau) &= j^{4/3} (\tau) - 992\ j^{1/3}(\tau)
\cr &= q^{-4/3} + 139504\, q^{2/3} + 69332992\, q^{5/3} + 6998296696\,
q^{8/3} \cr & \quad + 330022830080\, q^{11/3} + \ldots\ ,
}}
and
\eqn\zforty{\eqalign{Z_{40} (\tau) &= j^{5/3} (\tau) - 1240\, j^{2/3}(\tau)
\cr &= q^{-5/3} + 20620\, q^{1/3} + 86666240\, q^{4/3} + 24243884350\,
q^{7/3} \cr & \quad + 2347780456448\, q^{10/3} + \ldots\ .
}}
These partition functions have been first obtained by H\"ohn in
\Hoehn\ and the corresponding CFTs have been identified with
$\bbZ_2$ orbifolds of theories defined on even unimodular lattices
of the respective rank possessing no vectors of squared length 2.
Proceeding in this manner, we may in principle specify the
partition function for any value of $m$. Explicit formulas up to
$c=88$ (omitting the cases $c=48, \, 72$ already considered in
\WittenKT) are given below.
\eqn\zfiftysix{\eqalign{Z_{56} (\tau) &= j^{7/3} (\tau) - 1736\, j^{4/3}(\tau) + 401661\, j^{1/3}(\tau)
\cr &= q^{-7/3} + q^{-1/3} + 7402776\, q^{2/3} + 33941442214\, q^{5/3}
\cr & \quad + 16987600857280\, q^{8/3} + 2998621352249926\, q^{11/3} \ldots \ ,
\cr Z_{64} (\tau) &= j^{8/3} (\tau) - 1984\, j^{5/3}(\tau) + 705057\, j^{2/3}(\tau)
\cr &= q^{-8/3} + q^{-2/3} + 278512\, q^{1/3} + 13996663144\, q^{4/3} + 19414403055040\, q^{7/3}
\cr & \quad + 769385603725340\, q^{10/3} + 1062805058989221728\, q^{13/3} + \ldots\ ,
\cr
Z_{80} (\tau) &= j^{10/3} (\tau) - 2480\, j^{7/3}(\tau) +
1496361\, j^{4/3}(\tau) - 132423391\, j^{1/3}(\tau)
\cr &= q^{-10/3} + q^{-4/3} + q^{-1/3} + 173492852\, q^{2/3} + 4695630250012\, q^{5/3}
\cr & \quad + 8461738959649848\, q^{8/3} + 4293890043969667206\, q^{11/3} + \ldots\ ,
\cr
Z_{88} (\tau) &= j^{11/3} (\tau) - 2728\, j^{8/3}(\tau) +
1984269\, j^{5/3}(\tau) - 302198519\, j^{2/3}(\tau)
\cr &= q^{-11/3} + q^{-5/3} + q^{-2/3} + 2365502\, q^{1/3} + 907649518712\, q^{4/3}
\cr & \quad + 4712143513485758\, q^{7/3} + 4723281033156413468\, q^{10/3} + \ldots\ .}
}
These partition functions have not been, to our knowledge,
previously identified in the literature. It would be interesting
to examine whether corresponding CFTs actually exist.

\subsec{Microstate counting}

In what follows, we will verify that the partition functions
constructed above can account for the degeneracy of the BTZ black
hole states. According to Witten's interpretation, the new states
appearing at each level are divided into primary states,
corresponding to black holes, and Virasoro descendants of
lower-lying primary states, corresponding to lower-mass black
holes dressed with boundary excitations. Therefore, the number of
microstates associated with black holes of a given mass is given
by the number of primaries at the corresponding level. The total
number of states $D(m,L_0)$ at a given eigenvalue $L_0=h - {m
\over 3}$ is read off from the relation
\eqn\zexp{Z_{8m}(\tau) =
\sum_{L_0+m/3=0}^\infty D ( m,L_0 ) q^{L_0}\ ,
}
and the number $d(m,L_0)$ of primaries is then obtained by
subtracting the number of descendants at this level. Once
$d(m,L_0)$ is determined, we can define the microscopic entropy
\eqn\Sm{
S (m,L_0) = \ln d(m,L_0)\ .
}
In practice, the contribution of descendant states to $D(m,L_0)$
is negligible and we can trade $d(m,L_0)$ for $D(m,L_0)$. In any
case, the entropy computed by means of \Sm\ turns out to be quite
close to the semiclassical entropy \SBH, as explicitly shown on
Table 1 for $m=3,\ldots,8$ and for the first few values of $L_0$.

However, this agreement is a mild check of the proposed duality
since it is mostly controlled by modular invariance\foot{We thank
Edward Witten for stressing this point.} rather than by the
detailed structure of the theory. To see this, we recall that the
Petersson-Rademacher formula \refs{\Petersson,\Rademacher}
completely determines the coefficients $F(l)$ of $q^{l-c/24}$ in
the expansion of a modular form of weight $w$ in terms of the
corresponding polar coefficients $F(n)$, $n-{c \over 24}<0$,
according to \DijkgraafFQ
\eqn\rademacher{\eqalign{F(l) &= 2 \pi \sum_{n- c / 24 < 0}
\left( { {c \over 24} - n \over l - {c \over 24} }
\right)^{(1-w)/2} F(n)
\cr &\times \sum_{k=1}^\infty {1 \over k} \Klsum \left( l - {c \over 24},n - {c \over 24};k \right)
I_{1-w} \left( {4\pi \over k} \sqrt{ \left( {c \over 24} - n
\right) \left( l - {c \over 24} \right)} \right)\ , }
}
where $\Klsum (a,b\, ;k)$ is the Kloosterman sum
\eqn\kloosterman{\Klsum (a, b\, ;k) \equiv \sum_{d \in (\bbZ / k
\bbZ)^*} \exp \left( {2\pi \ii \over k} \left(d\,a + d^{-1}\, b
\right) \right)\ ,
}
and $I_\n(z)$ is a modified Bessel function of the first kind.
Using this expression for the coefficients $D(m,L_0)$ of the
partition function $Z_{8m}(\tau)$ and noting that for an extremal
CFT the polar coefficients are the same as those in the vacuum
character \znaive, namely $D\left(m,n-{m \over 3}\right) = P(n) -
P(n-1)$ for $n = 0,\ldots,\left[ m/3 \right]$, we find
\eqn\radextremal{\eqalign{D ( m,L_0 ) &= 2 \pi \sum_{n = 0}^{\left[ m/3 \right]} \sqrt{ {{m \over
3}-n \over L_0}
} \left( P(n)- P(n-1) \right)
\cr &\times \sum_{k=1}^\infty {1 \over k} \Klsum \left( L_0,n-{m \over 3};k \right)
I_1 \left( {4\pi \over k} \sqrt{\left( {m \over 3} - n \right)
L_0} \right)\ ,}
}
where we note that the $n=1$ term vanishes since $P(1)=P(0)$. In
this expression, the only factor that depends on the details of
the theory is $P(n)- P(n-1)$.
\topinsert
\parasize=1in
\noncenteredtables \line{
\begintable
$m$ | $L_0$ | $d$ | $S$ | $S_{BH}$
\crthick
    | $1$ | 196883 | 12.1904 | 12.5664
\nr
$3$ | $2$ | 21296876 | 16.8741 | 17.7715
\nr
    | $3$ | 842609326 | 20.5520 | 21.7656
\cr
    | $2/3$ | 139503 | 11.8458 | 11.8477
\nr
$4$ | $5/3$ | 69193488 | 18.0524 | 18.7328
\nr
    | $8/3$ | 6928824200 | 22.6589 | 23.6954
\cr
    | $1/3$ | 20619 | 9.9340 | 9.3664
\nr
$5$ | $4/3$ | 86645620 | 18.2773 | 18.7328
\nr
    | $7/3$ | 24157197490 | 23.9078 | 24.7812
\endtable
\hfil
\begintable
$m$ | $L_0$ | $d$ | $S$ | $S_{BH}$
\crthick
    | $1$ | 42987519 | 17.5764 | 17.7715
\nr
$6$ | $2$ | 40448921875 | 24.4233 | 25.1327
\nr
    | $3$ | 8463511703277 | 29.7668 | 30.7812
\cr
    | $2/3$ | 7402775 | 15.8174 | 15.6730
\nr
$7$ | $5/3$ | 33934039437 | 24.2477 | 24.7812
\nr
    | $8/3$ | 16953652012291 | 30.4615 | 31.3460
\cr
    | $1/3$ | 278511 | 12.5372 | 11.8477
\nr
$8$ | $4/3$ | 13996384631 | 23.3621 | 23.6954
\nr
    | $7/3$ | 19400406113385 |30.5963 | 31.3460
\endtable
}
\bigskip
\noindent Table 1: Degeneracies, microscopic entropies and
semiclassical entropies for the first few values of $m$ and $L_0$.
\endinsert
Eq. \radextremal\ is an exact result, which can be used to derive
various approximate expressions, appropriate for limiting cases. A
first simplification is to use Weil's estimate $\Klsum (a,b\, ;k)
\simeq \sqrt{k}$ to obtain the expression
\eqn\radweil{D ( m,L_0 ) \simeq 2 \pi \sum_{n = 0}^{\left[ m/3 \right]} \sqrt{ {{m \over
3}-n \over L_0} } \left( P(n)- P(n-1) \right) \sum_{k=1}^\infty {1
\over \sqrt{k}} I_1 \left( {4\pi \over k} \sqrt{\left( {m \over 3}
- n \right) L_0} \right)\ ,
}
which turns out to be in excellent agreement with the actual
number of microstates. The semiclassical results usually quoted in
the literature are obtained by taking the large-$m$ and
large-$L_0$ limit, using the asymptotics $I_1(z) \simeq e^z /
\sqrt{2\pi z}$, and keeping only the $n=0$ and $k=1$ terms in the
two summations. Doing so, all information about the details of the
theory (apart from the ground-state degeneracy) disappears, and
one obtains the Hardy-Ramanujan formula
\eqn\radhardy{D ( m,L_0 ) \simeq {1 \over \sqrt{2}} { (m / 3)^{1/4} \over L_0^{3/4} } \exp
\left( 4\pi \sqrt{{m \over 3} L_0 } \right)\ ,
}
which is valid for any CFT of central charge $c=8m$ and leads to
the entropy
\eqn\enthardy{S( m,L_0 ) \simeq S_{BH}( m,L_0 ) - {3 \over 2} \ln S_{BH}( m,L_0 ) + \ln {m \over 3} + \ln{4 \sqrt{2} \over \pi}\ ,
}
which is the Bekenstein-Hawking result plus the logarithmic
corrections \refs{\BirminghamXD,\KaulKF}. Therefore, in the
semiclassical limit one recovers the Bekenstein-Hawking entropy,
as guaranteed by Cardy's formula \CardyIE, while the qualitative
agreement for smaller values of $m$ and $L_0$, as those shown on
Table 1, is due to the convergence properties of the sums in
\radweil.

On the other hand, the exact formula \radextremal\ (or its
approximation \radweil) allows for a controlled expansion that
makes it possible to determine the various corrections to the
semiclassical results. In particular we note that the partition
numbers $P(n)$, being the coefficients of $q^{n-1/24}$ in
$\eta^{-1}(q)$, admit themselves the Petersson-Rademacher
expansion
\eqn\radpartition{P ( n ) = 2 \pi \left( {{1 \over 24} \over n - {1 \over 24}} \right)^{3/4}
\sum_{k=1}^\infty {1 \over k} \Klsum \left( n - {1 \over 24},- {1
\over 24};k \right) I_{3/2} \left( {4\pi \over k} \sqrt{ {1 \over
24} \left(n-{1 \over 24} \right)} \right)\ .
}
Substituting \radpartition\ (or a suitable approximation thereof)
in \radweil, and expanding around the semiclassical limit ($m,L_0
\to \infty$, $L_0/m$ fixed), one may in principle calculate all
leading and subleading corrections to the Bekenstein-Hawking
formula in a systematic manner. It has been reported in \WittenKT\
that a study of these corrections is in progress.

\newsec{Genus two extremal partition functions for $c=8m$}

The above construction of extremal partition functions can be
extended to the genus two case. In this case, the modular group is
$PSp(4,\bbZ)$ and the partition function is expressed as a
combination of Siegel modular forms (see e.g.
\refs{\Freitag,\Klingen}) with the appropriate transformation
properties. To determine the various coefficients, there are two
alternative methods. The first method \TuiteID\ is to consider the
limit in which the genus two surface degenerates to two tori
joined at a point\foot{Alternatively, one can consider the
degeneration to a single torus with two points joined. Consistency
requires that the two approaches be equivalent.} and to require
that the partition function factorize into a sum of products
involving two torus one-point functions and a certain power of the
pinching parameter; this method is quite straightforward but for
large $c$ it requires detailed information about the behavior of
Siegel forms under degeneration, which might be hard to obtain.
The second method, suggested in \WittenKT\ and used in \GaiottoXH\
for calculating the genus two partition functions for
$c=24,\,48,\,72$, is to determine the partition function from the
singularities of a six-point function of twist fields, possibly
using some restricted information from the factorization
condition; this method seems to be more involved than the first,
but is actually easier to apply for large values of $c$. In what
follows, we will use the first method to compute the genus two
partition functions for the extremal CFTs with $c=8m$ up to
$c=40$.

Before we do so, let us summarize the relevant formalism. In the
genus two case, the moduli space is the quotient of the Siegel
upper half space by the modular group $PSp(4,\bbZ)$, and is
parameterized by the period matrix $\Om$ which transforms
according to $\Om \to (A \Om + B)(C \Om + D)^{-1}$. A Siegel
modular form $F_w$ of weight $w$ is defined as a holomorphic
function of $\Om$ that transforms as follows
\eqn\modularsp{F_w(\Om) \to \left( \det(C \Om + D) \right)^w F_w(\Om)\ .
}
Any such form admits a Laurent expansion in the parameters
$q=e^{2\pi \ii \Om_{11}}$, $r=e^{2\pi \ii \Om_{12}}$ and
$s=e^{2\pi \ii \Om_{22}}$ or, equivalently, in $q$, $s$ and
$u=r+r^{-1}-2$. The ring of Siegel modular forms is generated
\Igusa\ by four forms of weight $4$, $6$, $10$ and $12$, namely
the two Eisenstein series $\psi_4$ and $\psi_6$ and the two cusp
forms $\chi_{10}$ and $\chi_{12}$. The former two admit the
following expansions
\eqn\eisen{\eqalign{\psi_4 (\Om) &= {1 \over 4} E_4(q) E_4(s) + 3600\, q s u + 60\, q s u^2 + \cO(q^2,s^2) \ ,
\cr \psi_6 (\Om) &= {1 \over 16} E_6(q) E_6(s) + 2646\, q s u + {63 \over 2} q s u^2 + \cO(q^2,s^2)\ ,}
}
where $E_4(q)=1 + 240\, q + \cO(q^2)$ and $E_6(q)=1 - 504\, q +
\cO(q^2)$ are the standard Eisenstein series, while the latter two
can be expanded in a similar manner with the leading terms given
in terms of the modular discriminant $\D(q)$ by $u \D(q) \D(s)$
and $96 \D(q) \D(s)$, respectively. To discuss the degeneration
limit, we will use the formalism of \refs{\TuiteID,\MasonDK}. The
genus two surface can be constructed according to a ``sewing''
procedure \SonodaMF, where two tori with modular parameters
$q_{1,2}= e^{2 \pi \ii \tau_{1,2}}$ are joined by excising a disc
of radius $|\e|$ from each torus ($\e$ being is a complex
``pinching'' parameter) and making an appropriate identification
of two annular regions around the excised discs. The degeneration
limit corresponds to $\e \to 0$, and the relations between the
parameters $q$, $r$, $u$ and $q_1$, $q_2$, $\e$ are as follows
\eqn\degen{\eqalign{q &=q_1 \left(1 - {1 \over 12} \e^2 E_2(q_2) \right) + \cO(\e^4) \ ,
\cr s &=q_2 \left(1 - {1 \over 12} \e^2 E_2(q_1) \right) + \cO(\e^4)\ ,
\cr u &= \e^2 + {1 \over 2} \e^4 \left( 1 + {1 \over 36} E_2(q_1) E_2(q_2) \right) + \cO(\e^6)\ ,}
}
where $E_2(q)=1 - 24\, q + \cO(q^2)$. The behavior of the Siegel
modular forms under degeneration is determined by substituting Eq.
\degen\ into the corresponding expansions. Doing so for
$\psi_{4,6}$ and applying the Ramanujan identities, we find
\eqn\eisendeg{\eqalign{\!\!\!\psi_4 &= {1 \over 4} E_{4,1} E_{4,2} + \e^2 \left( \! 3600\, q_1 q_2
+ {1 \over 144} ( E_{6,1} F_{6,2} + F_{6,1} E_{6,2} - 2 F_{6,1}
F_{6,2} ) \! \right) + \cO(\e^4)\ ,
\cr \!\!\!\psi_6 &= {1 \over 16} E_{6,1} E_{6,2} + \e^2 \left(\! 2646\, q_1
q_2 + {1 \over 384} ( E_{4,1}^2 F_{8,2} + F_{8,1} E_{4,2}^2 - 2
F_{8,1} F_{8,2} ) \!\right) + \cO(\e^4)\ , }
}
where $F_6 \equiv E_2 E_4$, $F_8 \equiv E_2 E_6$, while $E_{4,i}
\equiv E_4(q_i)$ and so on. Expanding the $\e^2$ term up to order
$q_i^2$, we obtain
\eqn\psideg{\eqalign{\!\!\!\!\psi_4 &= {1 \over 4} E_{4,1} E_{4,2} - 5\, \e^2 \Bigl( q_1+q_2 +  \left( 18(q_1^2 + q_2^2) - 288\, q_1 q_2 \right) + 216\, q_1 q_2 (q_1 + q_2)
\cr & \qquad\qquad\qquad\qquad\quad - 132192\, q_1^2 q_2^2 + \cO(q_i^3) \Bigr) +
\cO(\e^4)\ ,
\cr \!\!\!\!\psi_6 &= {1 \over 16} E_{6,1} E_{6,2} + {21 \over 8} \e^2 \Bigl( q_1+q_2 +  \left( 66(q_1^2 + q_2^2) - 48\, q_1 q_2 \right) - 39456\, q_1 q_2
(q_1 + q_2)
\cr & \qquad\qquad\qquad\qquad\qquad - 608256\, q_1^2 q_2^2 +
\cO(q_i^3) \Bigr) + \cO(\e^4)\ . }
}
Likewise, for $\chi_{10,12}$ we may use the formulas of
\refs{\TuiteID,\MasonDK}\foot{The weight-$12$ form $F_{12}$
appearing in \TuiteID\ is given by $F_{12} = {1 \over 9} ( 704\,
\psi_4^3 - 512\, \psi_6^2 + 76032\, \chi_{12} )$.} to find
\eqn\chideg{\eqalign{\!\!\chi_{10} &= \e^2 \left(\D_1 \D_2 - {1 \over 12} \e^2 \Bigl( q_1
q_2 - 48\, q_1 q_2 (q_1 + q_2) + 2304\, q_1^2 q_2^2 + \cO(q_i^3)
\Bigr) + \cO(\e^4) \right)\ ,
\cr \!\!\chi_{12} &= 96 \D_1 \D_2 - 8\, \e^2 \Bigl( q_1 q_2
+{9225 \over 22}q_1 q_2 (q_1 + q_2) + {1252764 \over 11} q_1^2
q_2^2 + \cO(q_i^3) \Bigr) + \cO(\e^4)\ ,}
}
where again $\D_i \equiv \D(q_i)$.

Returning to the problem at hand, the general form of the genus
two partition functions for the extremal CFTs under consideration
is dictated by $PSp(4,\bbZ)$ modular invariance. Namely, the genus
two partition function $Z^{(2)}_{8m} (\Om)$ for $c=8m$ should be
invariant up to a phase under the translations $T_1 \, : \,
\Omega_{11} \to \Omega_{11}+1$, $T_2 \, : \, \Omega_{22} \to
\Omega_{22}+1$, $U \, : \, \Omega_{12} \to \Omega_{12}+1$ and the
reflection $V \, : \, \e \to - \e$, and should satisfy the
transformation law $Z^{(2)}_{8m} (\Om) \to (-\tau_{1,2})^{2m/3}
Z^{(2)}_{8m} (\Om)$ under $S_{1,2} \, : \, \tau_{1,2} \to
-1/\tau_{1,2}\ , \e \to -\e/\tau_{1,2}$ \TuiteID. These conditions
lead us to expect that this function has the form
\eqn\zansatz{ Z^{(2)}_{8m} (\Om) = \chi_{10}^{- m / 3} (\Om)
\sum_{r=0}^{\left[ 2 m / 5 \right]} \chi_{10}^{r}(\Om)
\Psi_{4m-10r} (\Om)\ ,
}
where $\Psi_{4m-10r} (\Om)$ is an entire $PSp(4,\bbZ)$ modular
form of the indicated weight, constructed out of $\psi_4$,
$\psi_6$ and $\chi_{12}$. At the degeneration limit, this must
factorize as follows
\eqn\zfactor{
Z^{(2)}_{8m} (\Om) = \sum_h \e^{2h - 2m/3} \sum_{\phi_i,\phi_j \in
\cH_h} \cG^{\phi_i \phi_j} Z^{(1)}_{8m} ( \phi_i,\tau_1)
Z^{(1)}_{8m} ( \phi_j,\tau_2)\ ,
}
where the sums run over all levels $h$ and over all operators at
each corresponding subspace $\cH_h$ of the Hilbert space,
$Z^{(1)}_{8m} ( \phi_i,\tau)$ denotes the torus one-point function
of the operator $\phi_i$ (normalized so that $Z^{(1)}_{8m} (
1,\tau)$ equals the torus partition function
$Z^{(1)}_{8m}(\tau)$), and $\cG^{\phi_i \phi_j}$ is the inverse
Zamolodchikov metric (normalized so that $\cG^{1 1}=1$). By
inserting \psideg\ and \chideg\ in \zansatz\ and expanding in
$\e$, one may then determine each $\Psi_{4m-10r}$ by comparing the
$\cO(\e^{2r-2m/3})$ terms with the corresponding terms in
\zfactor. For $r=0$ this matching only requires knowledge of the
leading terms in \psideg\ and \chideg\ and of the torus partition
functions, while for each $r>0$ this matching requires knowledge
of the $\cO(\e^{2r})$ terms in \psideg\ and \chideg\ and of the
torus one-point functions of all operators at level $r$.
Furthermore, recalling that in the extremal CFTs of interest the
primaries appear at levels $h \geqslant \left[ m / 3 \right]+1$,
and noting that the torus one-point function of a primary with $h
\geqslant 11$ vanishes \GaiottoXH, we see that for $m \leqslant
32$ we may determine all $\Psi_{4m-10r}$ by knowing just the
one-point functions of the Virasoro descendants which are easy to
obtain using Ward identities or, equivalently, the more formal
methods of \Zhu.

For $m=1,2,3$ we can only have $r=0$, and the partition function
is completely determined by matching the $\cO(\e^{-2m/3})$ terms.
Doing so, we easily find
\eqn\zcsmall{\eqalign{Z^{(2)}_{8}(\Om) &= 4\, \chi_{10}^{- 1/3} \psi_4\ ,
\cr Z^{(2)}_{16}(\Om) &= 16\, \chi_{10}^{- 2/3} \psi_4^2\ ,
\cr Z^{(2)}_{24}(\Om) &= \chi_{10}^{- 1} \left( {328 \over 9} \psi_4^3 + {992 \over 9} \psi_6^2 - 7626\, \chi_{12} \right)\ ,}
}
with the expression in the last line being the same as that in
\refs{\TuiteID,\GaiottoXH}. As a consistency check, one may
compute the $\cO(\e^{2-2m/3})$ terms in \zansatz\ and \zfactor\
and verify that they match as well.

For $m=4,5$ we can have $r=0,1$, and determining the partition
function requires that we match the order $\cO(\e^{-2m/3})$ and
$\cO(\e^{2-2m/3})$ terms. To see how this matching works, let us
examine the case $m=4$ in detail. For this case, the ansatz
\zansatz\ takes the form
\eqn\zgtwoans{Z^{(2)}_{32} (\Om) = \chi_{10}^{- 4 / 3}
\left( A \psi_4^4 +  B \psi_4 \psi_6^2 + C \psi_4 \chi_{12} + D
\chi_{10} \psi_{6} \right)\ ,
}
where $A,\, B,\, C,\, D$ are coefficients to be determined.
Inserting the expansions \psideg\ and \chideg\ of the Siegel
modular forms, noting that the leading $\e^{-8/3}$ term depends
only on the combinations $E_{4,i}^3 / \D_i$ and $E_{6,i}^2 /
\D_i$, and using the identities
\eqn\jed{
j(\tau) = {E_4^3(q) \over \D(q)} = 1728 + {E_{6}^2(q) \over
\D(q)}\ ,
}
we can express the result as follows
\eqn\zgtwoansa{\eqalign{Z^{(2)}_{32} (\Om) &= {(j_1 j_2)^{1/3} \over \e^{8/3}}
\left( {4A+B \over 1024} j_1 j_2 - {27 B \over 16} (j_1+j_2) +
12(243B+2C) \right)
\cr &+ {(q_1 q_2)^{-1/3} \over \e^{2/3}} \biggl( {4A+B \over 9216} {1 \over q_1 q_2} + {124A - 23B \over 1152}
\left( {1 \over q_1} + {1 \over q_2} \right)
\cr &\qquad\qquad\qquad\quad + {3844A + 9277B + 96C + 9D \over 144} + \cO(q_i) \biggr) + \cO(\e^{4/3})\ , }
}
where $j_i \equiv j(\tau_i)$. Turning to the factorized expression
\zfactor, the $\cO(\e^{-8/3})$ and $\cO(\e^{-2/3})$ contributions
are due to the torus partition function and to the one-point
function of the Virasoro descendant $\phi$ at level two,
respectively. For this descendant field, the one-point function
equals $Z^{(1)}_{32}(\phi,\tau) = q {d \over dq}
Z^{(1)}_{32}(\tau)$ \DongEA\ while the Zamolodchikov metric is
$\cG_{\phi\phi} = {c \over 2} = 4 m = 16$. Hence, the relevant
terms in \zfactor\ read
\eqn\zgtwofac{
Z^{(2)}_{32} (\Om) = \e^{-8/3} \left( Z^{(1)}_{32} (\tau_1)
Z^{(1)}_{32} (\tau_2) + {1 \over 16} \e^2 \, q_1 {d
Z^{(1)}_{32}(\tau_1) \over dq_1}
 q_2 {d Z^{(1)}_{32}(\tau_2) \over dq_2} + \cO(\e^4)
\right)\ .
}
Inserting the explicit expression \zthirtytwo\ for
$Z^{(1)}_{32}(\tau)$ and expanding the second term in the $q_i$,
we obtain
\eqn\zgtwofaca{\eqalign{Z^{(2)}_{32} (\Om) &= {(j_1 j_2)^{1/3} \over \e^{8/3}}
\left( j_1 j_2 - 992 (j_1+j_2) + 984064 \right)
\cr &+ {(q_1 q_2)^{-1/3} \over \e^{2/3}} \biggl( {1 \over 9\, q_1 q_2} + \cO(q_i) \biggr) + \cO(\e^{4/3})\ . }
}
In order for the $\cO(\e^{-8/3})$ terms in \zgtwoansa\ and
\zgtwofaca\ to match for all values of $q_1$ and $q_2$, the
coefficients of the various powers of $j_1$ and $j_2$ in the two
expressions should be equal. This fixes the coefficients $A,B,C$
to the values $A={2944 \over 27}$, $B={15872 \over 27}$ and
$C=-{91264 \over 3}$. Inserting these values in the
$\cO(\e^{-2/3})$ terms in \zgtwoansa\ we verify the remarkable
fact that the terms of order $q_i^{-4/3}$ automatically match,
while the requirement that the terms of order $(q_1 q_2)^{-1/3}$
match as well fixes $D=-{984064 \over 3}$. Therefore, our final
result for the genus two partition function reads
\eqn\zgtwothirtytwo{Z^{(2)}_{32} (\Om) = \chi_{10}^{- 4 / 3}
\left( {2944 \over 27} \psi_4^4 + {15872 \over 27} \psi_4 \psi_6^2
-{91264 \over 3} \psi_4 \chi_{12} -{984064 \over 3} \chi_{10}
\psi_{6} \right)\ .
}
Repeating the same steps for $m=5$, we find that the corresponding
genus two partition function is given by
\eqn\zgtwoforty{Z^{(2)}_{40} (\Om) = \chi_{10}^{- 5 / 3}
\left( {7808 \over 27} \psi_4^5 + {79360 \over 27} \psi_4^2
\psi_6^2 - {302560 \over 3} \psi_4^2 \chi_{12} - {9840640 \over 3}
\chi_{10} \psi_4 \psi_{6} \right)\ .
}
The procedure can be extended to higher values of $m$, but becomes
increasingly cumbersome since one needs higher-order terms in the
$\e$- and $q_i$-expansions of Siegel modular forms in Eqs.
\psideg\ and \chideg. For example, for $m=6$ the expansions given
here allow us to determine the polynomials $\Psi_{24}$ and
$\Psi_{14}$, in precise agreement with the results of \GaiottoXH,
but leave the coefficient in $\Psi_4 \sim \psi_4$ undetermined.
For such values of $m$, the method proposed in \WittenKT\ is
perhaps more appropriate.

To summarize this section, we have applied the method of sewing
tori to compute the genus two partition functions of the extremal
CFTs up to $c=40$, providing thus a non-trivial consistency check
of these theories. It would be interesting to verify the above
expressions by considering the alternative degeneration limit to a
single torus with the aid of the formulas in \MasonDK. It is also
important to try to construct genus two partition functions for
larger values of $c$, where it is not known whether extremal CFTs
actually exist.

\newsec{Discussion}

In this paper, we contemplated the possibility that the
conjectured duality of \WittenKT\ between three-dimensional ${\rm
AdS}$ gravity and extremal holomorphic CFTs at central charges
that are multiples of 24 may actually apply to the more general
case of central charges that are multiples of 8. Although for
these cases holomorphic factorization is not possible, one can
impose holomorphic factorization up to a phase and follow the
reasoning of \WittenKT\ to uniquely determine the partition
functions of the CFTs with the required properties. Here, we
explicitly computed the genus one partition functions of these
theories up to $c=88$, we gave general expressions for determining
the degeneracies of states, and we also calculated the genus two
partition functions of the theories up to $c=40$.

Certainly, there is a number of open issues with the CFTs under
consideration and with their conjectured relation to
three-dimensional gravity. Regarding the CFTs, the fact that the
existence and uniqueness of of the $c \geqslant 48$ theories has
not been established requires performing further consistency
checks that go beyond modular invariance; in fact, there are
indications \GaberdielVE\ that the theories with $c=24k$ and $k
\geqslant 42$ may be inconsistent. Regarding the proposed duality,
we are lacking a solid argument in favor of holomorphic
factorization and a rationale for including the resulting
non-geometric configurations (e.g. a black hole in the holomorphic
sector and ${\rm AdS}$ space in the antiholomorphic sector
\WittenKT) in the gravity path integral. Moreover, as the
agreement with the semiclassical entropy is by no means compelling
evidence for the duality, additional arguments in support of it
are needed. We hope that further developments will shed light on
these issues.

\bigbreak\bigskip\centerline{{\bf Acknowledgements}}\nobreak
\smallskip
\noindent The authors are grateful to Jorge Russo and Konstadinos
Sfetsos for discussions, and to Michael Tuite and Edward Witten
for helpful correspondence. This work is co-funded by the European
Social Fund (75\%) and by National Resources (25\%) $-$ (EPEAEK
II) $-$ PYTHAGORAS.

\listrefs

\end